\documentstyle[seceq,epsf,wrapft,twoside]{ptptex}
\newcommand{\etal}{\it et al. \rm}
\newcommand{\rt}{\rightarrow}
\newcommand{\pipi}{\pi^+ \pi^-}
\newcommand{\ppbar}{p \overline{p}}
\newcommand{\pbar}{\overline{p}}

\newcommand{\jpsi}{J/\psi}
\newcommand{\ee}{e^+ e^-}
\notypesetlogo  
\title{%
Observation of a narrow, near-threshold enhancement 
in the {\boldmath $\ppbar $} mass spectrum from 
radiative {\boldmath $\jpsi\rt\gamma\ppbar$} decays}
\author{
Stephen L.  {\sc Olsen} }
\inst{%
Department of Physics and Astronomy\\ 
The University of Hawaii at Manoa\\
2505 Correa Road, Honolulu, HI 96822, USA\\
(Representing the BES Collaboration)\\ }
\recdate{%
\today
}
\abst{%
We observe a narrow enhancement near $2m_p$ in the invariant 
mass spectrum of $\ppbar$ pairs from radiative $\jpsi\rt\gamma\ppbar$
decays.   No similar structure is seen in $\jpsi\rt \pi^0\ppbar$ decays.
The results are based on an analysis of a 58 million event 
sample of $\jpsi$ decays accumulated with the BESII detector at the
Beijing electron-positron collider.  The enhancement can be fit with
either an $S$- or $P$-wave Breit Wigner resonance function;
in the case of the $S$-wave fit, the peak mass is below $2m_p$ at
$M = 1859 ^{~+3}_{-10}~{\rm (stat)} ^{~+5}_{-25}~{\rm (sys)}~{\rm 
MeV}/c^2$ and the total width is $\Gamma < 30 $~MeV/$c^2$ at the 90 
percent confidence level. These mass and width values
are not consistent with the properties of any known particle.}
\begin{document}

\maketitle

\setcounter{tocdepth}{4}
\section{Introduction}
There is an accumulation of evidence for anomalous behavior in the
proton-antiproton ($\ppbar$) system very near the $M_{\ppbar}=2m_p$
mass threshold.
The observed cross section\cite{FENICE} for $\ee\rt hadrons$
has a narrow dip-like structure
at a center of mass energy of 
$\sqrt{s} \simeq 2m_p c^2$.
The proton's time-like magnetic
form-factor, determined from high statistics measurements of the
$\ppbar\rt\ee$ annihilation process,  exhibits a very steep fall-off
just above the $\ppbar$ mass threshold\cite{LEAR}.
The authors of ref.~1 attribute these features as being
due to the effects of a narrow, $S$-wave triplet 
$\ppbar$ resonance with $J^{PC} = 1^{--}$, a mass 
of 1870~MeV/$c^2$, and a width of 10~MeV/$c^2$.  
In studies of $\pbar$ annihilations
at rest in deuterium, anomalies in the charged pion momentum 
spectrum from $\pbar d\rt \pi^-\pi^0 p$ and $\pi^+\pi^- n$ 
reactions\cite{bridges} and the spectator proton spectrum from 
the $\pbar d\rt 2\pi^+ 3\pi^- p_s$ process\cite{dalkarov} have
been interpreted as effects of a narrow sub-threshold
resonance with properties 
similar to those of the proposed $1^{--}$ state listed above.
There are no well established mesons that could
be associated with such a state.  
Belle\cite{Belle1} has reported observations of the decays
$B^+\rt K^+\ppbar$ and
$\overline{B}{}^0\rt D^0\ppbar$. In 
both processes there are enhancements
in the ${\ppbar}$ invariant mass distributions
near $M_{\ppbar}\simeq 2m_p$.

The proximity in mass
to $2m_p$ is suggestive of nucleon-antinucleon ($N\overline{N}$) 
bound states,  an idea that has a long history.
In 1949, Fermi and Yang\cite{Fermi} proposed
that the pion was a tightly bound $N\overline{N}$ state.  
Although this turned out not to be correct, the train
of thought started by this paper had enormous consequences.
In 1956, after the discovery of the strange particles
$\Lambda$ and $K^0$, Sakata\cite{sakata} expanded this picture and used
``fundamental baryon triplets'' comprised of ($p, n, \Lambda$)
and their antiparticles ($\bar{p}, \bar{n}, \bar{\Lambda}$)
to make both pions and kaons.  It was subsequently 
realized that the underlying mathematics of this model 
was that of the $SU(3)$ unitary group\cite{ohnuki}, a  
realization that inevitably led to the quark model.  In 1961, 
Nambu and Jona-Lasinio\cite{Nambu} introduced
a dynamical theory based on chiral invariance that 
also considers mesons as baryon-antibaryon composites.
In this model, in addition to a low-mass pion, 
there is a scalar $\ppbar$ composite state with mass equal
to $2m_p$.  Here again, these ideas have been
superseded by the quark model\cite{kunihiro},
and the scalar composite state is the famous
$\sigma$ meson that is a main subject of discussion
at this meeting. 

Although in both examples cited above the original motivation 
for nucleon-antinucleon composites is gone, the possibility of
bound $N\overline{N}$ states with mass near $2m_p$, {\it i.e.,}
$p$-$\pbar$ analogues of the deuteron and
generally referred to as {\it baryonium}, continues to be 
considered\cite{richard}.  An investigation of low mass 
$\ppbar$ systems with different quantum numbers 
may help clarify the situation.

In this talk I report on a study of the low mass $\ppbar$ pairs
produced via radiative decays in a sample of 58 million $\jpsi$ 
events accumulated in the upgraded Beijing Spectrometer (BESII) 
located at the  Beijing Electron-Positron Collider (BEPC) at the 
Beijing Institute of High Energy Physics.  In this reaction,
the $\ppbar$ pair is produced in an environment that is
free of any other hadrons.  Moreover,
charge-parity conservation insures that 
the $\ppbar$ system has $C=+1$.

\section{Experimental considerations}

BESII is a large-solid-angle magnetic spectrometer 
that is described in detail elsewhere\cite{BES}.
For this analysis we  use events with a high energy gamma ray
and two oppositely charged tracks that are identified
as protons by their specific ionization ($dE/dx$) in
the tracking chamber.  Since antiprotons that stop
in the material of the detection systems can produce
annihilation products that are reconstructed
elsewhere as $\gamma$ rays, no restrictions are placed 
on the total number of photons in the event.

\begin{wrapfigure}{r}{5.5cm}
  \epsfysize=5.0cm
   \centerline{\epsffile{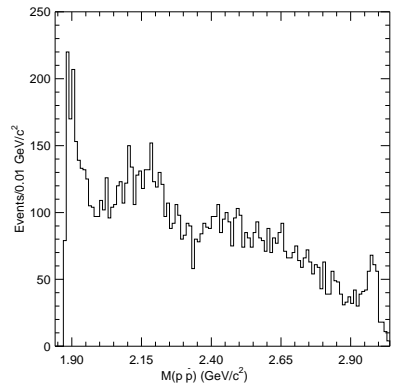}}
  \caption{
The $\ppbar$ invariant mass distribution for the
$\jpsi\rt\gamma \ppbar$-enriched event sample
 }
\label{fig:2pg_data_ihep}
\end{wrapfigure}

We subject the selected event candidates to four-constraint kinematic 
fits to the hypotheses $\jpsi\rt\gamma\ppbar$ and $\jpsi\rt\gamma 
K^+K^-$. For events with more that one $\gamma$, we 
select the $\gamma$ that has the highest fit confidence level.
We select events that have fit confidence level 
$CL_{\gamma\ppbar}>0.05$ and reject events that have
$CL_{\gamma K^+K^-} > CL_{\gamma\ppbar}$.

Figure~\ref{fig:2pg_data_ihep} shows the $\ppbar$ invariant mass
distribution for surviving events.  The distribution has a peak
near $M_{\ppbar}=2.98$~GeV/$c^2$ that is consistent in mass, width, 
and yield with expectations for $\jpsi\rt\gamma\eta_c$,
$\eta_c\rt\ppbar$\cite{dong},  a broad enhancement around 
$M_{\ppbar}\sim 2.2$~GeV/$c^2$, and a narrow, low-mass peak at the 
$\ppbar$  mass threshold that is the main subject of this talk\cite{gppb_prl}.

\subsection{Backgrounds}

Backgrounds from processes involving charged particles that are
not protons and antiprotons are negligibly small.
In addition to being well separated from other charged particles
by the $dE/dx$ measurements and the kinematic fit, the protons and 
antiprotons from the low $M_{\ppbar}$ region tend to stop in the
material in front of the electromagnetic shower detector,
where they have very characteristic responses: protons do not
produce any signals in the shower detector while secondary
particles from antiproton annihilation usually produce
large signals. This asymmetric behavior is quite distinct from that
for $K^+ K^-$, $\pipi$ or $\ee$ pairs, where the positive and negative 
tracks produce similar, non-zero responses.  The observed shower 
counter energy distributions for the low-mass $\jpsi\rt\gamma\ppbar$ events 
closely match expectations for protons and antiprotons and show no
evidence for contamination from other particle species.

\begin{wrapfigure}{r}{6.5cm}
  \epsfysize=6cm
   \centerline{\epsffile{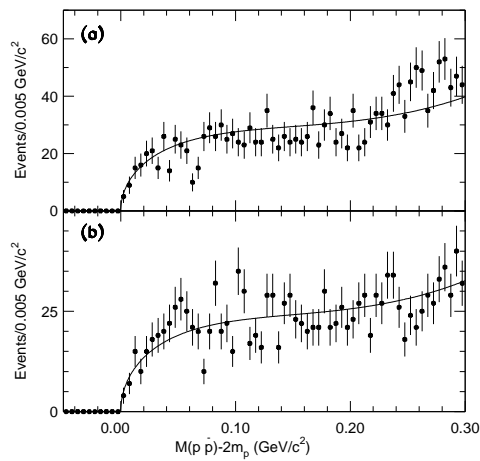}}
  \caption{
The $M_{\ppbar}-2m_p$ distribution for {\bf (a)}
selected $\jpsi\rt\pi^0 \ppbar$ decays and
{\bf (b)} MC $\jpsi\rt\pi^0 \ppbar$ events that satisfy
the $\gamma\ppbar$ selection criteria. 
The smooth curves 
are the result of fits described in the text.
 }
\label{fig:ppbar_mass}
\end{wrapfigure}

There is, however, a large background from $\jpsi\rt\pi^0\ppbar$ events
with an asymmetric $\pi^0\rt\gamma\gamma$ decay where
one of the photons has most of the $\pi^0$ energy.  
This is studied using a sample of $\jpsi\rt\pi^0\ppbar$
decays reconstructed from the same data sample.  For
these, we select
events with oppositely charged tracks that are identified as protons
and with two or more photons, 
apply a four-constraint kinematic fit
to the hypothesis $\jpsi\rt\gamma\gamma\ppbar$, and
require $CL_{\gamma\gamma\ppbar}>0.005$. For
events with more than two $\gamma$'s, we select the $\gamma$ pair that
produces the best fit.  In the $M_{\gamma\gamma}$ distribution
of the selected events there is a very
distinct $\pi^0$ signal; we require 
$\vert M_{\gamma\gamma}-M_{\pi^0}\vert < 0.03$~GeV/$c^2$ 
($\pm 2\sigma$).
The distribution of events {\it vs.} $M_{\ppbar} - 
2m_p$ near the $M_{\ppbar} = 2m_p$ threshold,
shown in Fig.~\ref{fig:ppbar_mass}(a), is reasonably
well described by a function of the form 
$f_{\rm bkg}(\delta) =N(\delta^{1/2}+ a_1 \delta^{3/2} + a_2
\delta^{5/2})$,
where $\delta \equiv M_{\ppbar} - 2m_p$ and the shape parameters
$a_1$ and $a_2$ are determined from a fit to simulated MC events
that were generated uniformly in phase space.   This is
shown in the figure as a smooth curve. There is no indication of a
narrow peak at low $\ppbar$ invariant masses. Monte Carlo simulations of
other $J/\psi$ decay processes with final-state $\ppbar$ pairs 
indicate that backgrounds from processes other than 
$J/\psi\rt\pi^0 \ppbar$ are negligibly small.

The $M_{\ppbar}-2m_p$ distribution for the $\pi^0\ppbar$ phase-space 
MC events that pass the $\gamma\ppbar$ selection
is shown in Fig.~\ref{fig:ppbar_mass}(b).
There is no  clustering at threshold;  the
smooth curve is the result of a fit to $f_{\rm bkg}(\delta)$ with 
the same shape parameter values.

\section{Results}

\begin{wrapfigure}{r}{6.5cm}
  \epsfysize=6cm
   \centerline{\epsffile{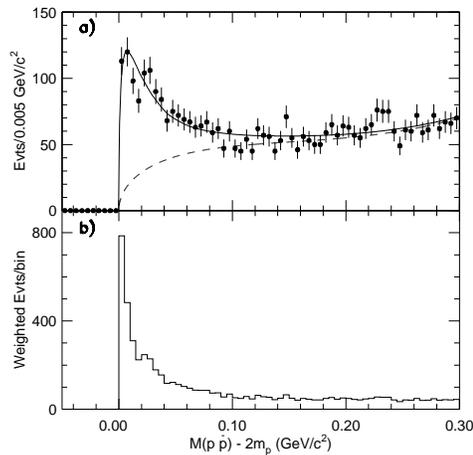}}
  \caption{
(a) The near-threshold $M_{\ppbar}-2m_p$ distribution for
the $\gamma\ppbar$ event sample. The solid curve   
is the result of the fit described  in the text;
the dashed curve shows the fitted background function.
(b) The  $M_{\ppbar}-2m_p$ distribution with events weighted
by $q_0/q$.
 }
\label{fig:2pg_thresh_fit}
\end{wrapfigure}

Figure~\ref{fig:2pg_thresh_fit}(a) shows the 
near-threshold $M_{\ppbar}-2m_p$ distribution 
for the selected $\jpsi\rt\gamma\ppbar$ events.  
The solid curve shows the result of a fit using an 
acceptance-weighted $S$-wave Breit-Wigner (BW)
function\cite{kcube} to represent the low-mass enhancement 
plus $f_{\rm bkg}(\delta)$ to represent the background.
The mass and width of the BW signal function are allowed 
to vary and the shape parameters of $f_{bkg}(\delta)$ are 
fixed at the values derived from the fit to the 
$\pi^0\ppbar$ phase-space MC sample\cite{resol}.  This fit 
yields $928\pm 57$ events in the BW function with a peak 
mass of $M=1859 ^{~+3}_{-10}$~MeV/$c^2$ and a full width 
of $\Gamma = 0^{+21}_{~-0}$~MeV/$c^2$.  Here the errors are 
statistical only.  The fit confidence level is 46.2\% 
($\chi^2/d.o.f.= 56.3/56$).

Further evidence that the peak mass is below the $2m_p$ 
threshold is provided in Fig.~\ref{fig:2pg_thresh_fit}(b), 
which shows the $M_{\ppbar}-2m_p$ distribution when the 
kinematic threshold behavior is removed by weighting each 
event by $q_0/q$, where $q$ is the proton momentum in the 
$\ppbar$ restframe and $q_0$ is the value for 
$M_{\ppbar}=2$~GeV/$c^2$.  The sharp and monotonic increase 
at threshold that is observed in this weighted histogram can 
only occur for an $S$-wave BW function when the peak mass 
is below $2m_p$.

An $S$-wave $\ppbar$ system with even $C$-parity 
would correspond to a $0^{-+}$ pseudoscalar state.
We also tried to fit the signal with a $P$-wave BW function,
which would correspond to a $0^{++}$ ($^3_0P$) scalar state 
that occurs in some models\cite{Nambu,richard}. 
This fit yields a peak mass $M=1876.4 \pm 0.9$~MeV$c^2$, 
which is very nearly equal to $2m_p$, and a very narrow total
width: $\Gamma = 4.6 \pm 1.8$~MeV$c^2$ (statistical errors only).  
The fit quality, $\chi^2/d.o.f. = 59.0/56$, is worse 
than that for the $S$-wave BW but still acceptable.
A fit with a $D$-wave BW fails badly with
$\chi^2/d.o.f. = 1405/56$.

\subsection{Can this be the effect of any known resonance?}
In addition we tried fits that use known particle resonances to 
represent the low-mass peak.  There are two spin-zero resonances 
listed in the PDG tables in this mass region\cite{PDG}: 
the $\eta(1760)$ with  $M_{\eta(1760)} = 1760 \pm 11$~MeV/$c^2$ and 
$\Gamma_{\eta(1760)} = 60 \pm 16$~MeV, and the $\pi(1800)$ with
$M_{\pi(1800)} = 1801 \pm 13$~MeV/$c^2$ and 
$\Gamma_{\pi(1800)} = 210 \pm 15$~MeV. A fit with $f_{\rm bkg}$ 
and an acceptance-weighted $S$-wave BW function with mass
and width fixed at the PDG values for the $\eta(1760)$ produces
$\chi^2/d.o.f. = 323.4/58$.  A fit using a BW with the 
$\pi(1800)$ parameters is worse.

\subsection{Production angle distribution}

\begin{wrapfigure}{r}{5.5cm}
  \epsfysize=5cm
   \centerline{\epsffile{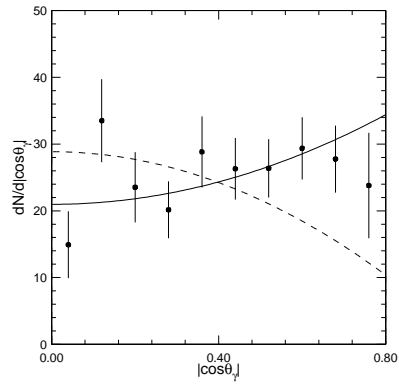}}
  \caption{
The background-subtracted, acceptance-corrected
$|\cos\theta_{\gamma} |$ distribution for
$\jpsi\rt\gamma\ppbar$-enriched events with $M_{\ppbar}\le
1.9$~GeV/$c^2$.  The solid curve is a fit to
a $1+\cos^2\theta_{\gamma}$ shape
for the region $|\cos\theta_{\gamma}|\le 0.8$;
the dashed curve is the result of a 
fit to $\sin^2\theta_{\gamma}$.
 }
\label{fig:2pg_cosg_corr}
\end{wrapfigure}

For both the scalar or pseudoscalar case, the  polar angle 
of the photon, $\theta_{\gamma}$, would be distributed
according to $1 + \cos^2\theta_{\gamma}$.
Figure~\ref{fig:2pg_cosg_corr} shows the
background-subtracted, acceptance-corrected 
$|\cos\theta_{\gamma}|$ distribution for events with
$M_{\ppbar}\le 1.9$~GeV and $|\cos\theta_{\gamma}|\le 0.8$.  
Here we have subtracted the $\vert \cos\theta_{\pi^0}\vert $
distribution from the $\pi^0\ppbar$ data sample, normalized to
the area of $f_{\rm bkg}(\delta)$ for $M_{\ppbar}<1.9$~GeV/$c^2$ to
account for background. The solid curve shows the result of a fit 
for $1 + \cos^2\theta_{\gamma}$ to the 
$\vert\cos\theta_{\gamma}\vert<0.8$ region; the dashed line shows 
the result of a similar fit to $\sin^2\theta_{\gamma}$.  
Although the data are not precise enough to establish a 
$1 + \cos^2\theta_{\gamma}$ behavior, the distribution is 
consistent with expectations for a radiative transition
to a pseudoscalar or scalar meson\cite{angle}.

\subsection{Systematic Errors}

We evaluate systematic errors on the mass and width
from changes observed in the fitted values for
fits with different bin sizes, with background shape 
parameters left as free parameters, different shapes
for the acceptance variation, and different resolutions.  
A study based on an ensemble of Monte Carlo experiments
for sub-threshold resonances demonstrates that, in the presence of
background, the farther the peak is below threshold,
the less reliable is the determination of its mass.  
The MC studies also indicate that the mass determination 
of a below-threshold resonance can be biased.   We include the
range of differences between input and output values seen
in the MC study in the systematic errors.

For the mass, we determine a systematic error 
of $^{~+5}_{-25}$~MeV$c^2$.  For the total width, we 
determine a 90\% CL upper limit of $\Gamma < 30$~MeV/$c^2$, 
where the limit includes the systematic error.

\subsection{Branching Fraction}

Using a MC-determined acceptance of $23\%$, we determine
a product of branching fractions ${\cal B}(\jpsi\rt\gamma X(1859))
{\cal B}(X(1859)\rt\ppbar) = (7.0 \pm 0.4 {\rm (stat)} 
^{+1.9}_{-0.8}{\rm (syst)})\times 10^{-5}$, where the systematic 
error includes uncertainties in the acceptance (10\%), the total 
number of $\jpsi$ decays in the data sample (5\%), and the effects 
of changing the various inputs to the fit ($^{+24\%}_{~-2\%}$).

\section{Summary}

In summary, we observe a strong, near-threshold enhancement in the
$\ppbar$ invariant mass distribution in the radiative decay
process $\jpsi\rt\gamma\ppbar$. No similar structure is seen 
in $\jpsi\rt \pi^0\ppbar$ decays.  The structure has properties 
consistent with either a $J^{PC} = 0^{-+}$ or $0^{++}$ quantum 
number assignment and cannot be attributed to the effects of any 
known meson resonance.  If interpreted as a single $0^{-+}$ resonance,
its peak mass is below the $M_{\ppbar}=2m_p$ threshold at
$1859 ^{~+3}_{-10} {\rm (stat)} ^{~+5}_{-25} {\rm (syst)}$ MeV/$c^2$ and
its width is $\Gamma < 30 $~MeV/$c^2$ at the 90\% CL.
These parameters are quite similar to those of the
$1^{--}$ state proposed in ref.~1, which strongly
suggests that these states may be related. 

\acknowledgements

The author would like to express his sincere gratitude to 
Professors S. Ishida  and K. Takamatsu and their colleagues
for inviting me to speak at this meeting and for their gracious 
hospitality.  I also thank Professors J.~Rosner and S.F.~Tuan for 
useful discussions.  The BES collaboration thanks the staffs of 
BEPC and the IHEP computing center for their strong efforts.


\begin{thebibliography}{99}
\bibitem{kcube} For the BW, we use the form:
\[
BW(M) \propto \frac{q^{(2\ell+1)} k^3}{(M^2-M_0^2)^2 + M_0^2\Gamma^2},
\]
where $M_0$ and $\Gamma$ are constants (determined from the fit), 
$q$ is the proton momentum in the $\ppbar$ restframe,
$\ell$ is the $\ppbar$ orbital angular momentum, and $k$ is the 
photon momentum.
\bibitem{resol} 
The $\ppbar$ invariant mass
resolution 
varies from $\sigma\simeq 1.2$~MeV/$c^2$
at $M_{\ppbar}\simeq 2m_p$,
to $\sim 3$~MeV/$c^2$ at higher
masses.  Convolving the fitting function with
a Gaussian with a width in this range
has no significant effect on the results.
\bibitem{angle} 
The $1+\cos^2\theta_{\gamma}$ fit has $\chi^2/d.o.f. = 3.7/7$; the
$\sin^2\theta_{\gamma}$ fit has $\chi^2/d.o.f. = 10.3/7$.
\end{thebibliography}
\end{document}